# N-Graphene Synthesized in Astrochemical Ices

K K Rahul[1], M Ambresh[2], D Sahu[3], J K Meka[1], S -L Chou[4], Y -J Wu[4], D Gupta[5], A Das[6], J -I Lo[4,7], B -M Cheng[4,7], B N Raja Sekhar[8], A Bhardwaj[1], H Hill[9], P Janardhan[1], N J Mason[10], B Sivaraman[1, *]

[1]Physical Research Laboratory, Navrangpura, Ahmedabad, India.

[2]Center for Nanoscience, Indian Institute of Science, Bangalore, India.

[3]Academia Sinica, Institute of Astronomy and Astrophysics, Taiwan.

[4]National Synchrotron Radiation Research Center, Hsinchu, Taiwan.

[5]University of Rennes-1, CNRS, IPR (Institute de Physique de Rennes), Rennes, France.

[6]Indian Centre for Space Physics, Kolkata, India.

[7]Department of Medical Research, Hualien Tzu Chi Hospital, Buddhist Tzu Chi Medical Foundation, Hualien, Taiwan.

[8]Atomic and Molecular Physics Division, Bhabha Atomic Research Centre, Mumbai, India.

[9]International Space University, Strasbourg, France.

[10]School of Physical Sciences, University of Kent, Canterbury, UK.

**Abstract**

Icy mantles of benzonitrile, an aromatic with a cyanide side chain that has recently been detected in the interstellar medium, were subjected to vacuum ultraviolet photon irradiation and found to form a residue. The residue was removed from the substrate and placed on a Quantifoil grid for electron microscopy analysis. Transmission electron microscopy showed Quantum Dot (QD) and Nitrogen-doped Graphene (N-Graphene) sheets. Diffraction and Energy Dispersive X-ray Spectroscopy revealed the crystalline nature and carbon-nitrogen composition, of the observed graphene sheet. This is the first result showing QD and N-Graphene synthesis in ice irradiation at interstellar temperatures.

\* Corresponding author – bhala@prl.res.in

1. Introduction

Cold ice covered dust particles in the InterStellar Medium (ISM) are known to be the tiny chemical factories responsible for the evolution of complex molecules observed in the ISM. The molecular complexity of the icy mantles on these cold dust grains is driven by energetic particles irradiation and temperature changes. Although the molecular ices present as icy mantles on a dust grain are volatile the change in physico-chemical nature brought by the irradiation can lead to nucleation [1] producing refractory residues that may be observed at higher temperatures when simpler species are sublimed from the surface. Spectral signatures of such a residue obtained from a mixture of organic and inorganic molecular ice [2] has been used to explain the 3.4 μm signature in ISM clouds from space based observations.

The nature of the residue depends on the chemical composition of the icy mantle undergoing irradiation under conditions commensurate to those found in the ISM and in the Solar System [3]. Most of the residues are complex organic [1,4-9] and inorganic [10-14] compounds. For cometary ices present in the Oort cloud, such residues may form an outer web of non-volatile coating [15,16] preventing further ice irradiation. The recent discovery of an ammoniated refractory residue [17,18] on comets may be a result of ice irradiation. An icy mixture containing simple ices, upon energetic processing, may also leave bio-residues [19] that might be of interest in the *Origin of Life* research.

There is an urgent need to understand the physical nature of the residue as it may play a key role [20] in the chemical evolution. Residue from hydrocarbon ice irradiation was reported [21] to be fluffy and porous due to polymerization on carbon-rich ices [22] leading to micrometer-sized polymer layers [23]. Nevertheless, the residues obtained by irradiating an aromatic (benzene) molecular ice were found to contain geometric shapes such as micron sized cubes and spheres, surrounded by nanoparticles [24]. The chemical analysis of the aromatic residues contained aromatic derivatives, from biphenyls to polyphenyls [25], characteristic spectral signatures of aromatic derivatives were present in the Vacuum UltraViolet (VUV) photoabsorption spectrum of the residue [24]. Two possibilities for the residue can be discussed, (i) Mixed Aromatic/aliphatic Organic Nanoparticles (MAONs) as dust in the place of (ii) Polycyclic Aromatic Hydrocarbons (PAH), based on space based observations in the infrared [26,27]. So there is a dire need for more laboratory experiments to understand the physico-chemical nature of the residue from astrochemical ice irradiation.

2. Search for graphene in the ISM

The report of the possible presence of $C_{24}$, a small graphene sheet [28,29] has led to estimates of the concentrations of graphene in different regions of the ISM [30] and suggest the presence of larger graphene sheets. Furthermore, it is expected that the presence of graphene may be concomitant with the presence of Buckminster fullerene, $C_{60}$, and the larger fullerene, $C_{70}$, already identified in the ISM [31]. The formation of $C_{60}$ is proposed to follow one of two chemical pathways; bottom – up [32,33] and top – down [34] model. The top-down model starts with PAH being converted to a larger graphene sheet which eventually synthesizes $C_{60}$ following the pathway suggested by Berne and Tielens [35]. Therefore, the presence of $C_{60}$ can be used as a tracer to look for graphene rich clouds in the ISM.

Perhaps, hydrogen loss from MAONs may also form graphene. However, to-date, the formation of these allotropes of carbon have only been examined at higher temperatures that do not resemble the cold dust containing an icy mantle. Efforts were taken to understand the formation of carbon clusters in icy conditions, up to $C_{20}$ were observed to form at low temperature in an irradiated methane in neon ice matrix [36]. The results reported were convincing that carbon atoms combine to form chain of molecules, even at such lower temperatures, which if further pursued may lead to the formation of graphene.

The recent identification of molecules containing an aromatic ring with a side chain in the ISM, such as benzonitrile [37] that can readily form [38] at ISM conditions, provides another possible route to the formation of complex carbon structures such as graphene. Irradiation of such a compound may lead to dissociation liberating the aromatic ring which may then assemble with other rings to form a more complex PAH. In this paper we report for the first time the results of photoprocessing of benzonitrile ice prepared under ISM conditions.

### 3. The laboratory analogue experiment and observed carbonaceous residues

VUV photoprocessing of benzonitrile ices was studied at two beamlines of Taiwan Light Source (TLS), TLS BL03 and TLS BL21A2, at the National Synchrotron Radiation Center. The details of the experimental system used can be found elsewhere so only salient details are presented here [36,39,40].

A benzonitrile ice film was formed at 4 K, by depositing benzonitrile at $5 \times 10^{-7}$ Torr for 60 seconds onto Lithium Fluoride (LiF) and Potassium Bromide (KBr) substrates. The ice was then irradiated by 9 eV photons for nearly 9 hours and subsequently monitored using VUV and InfraRed (IR) spectroscopy. The amount of sample deposited is quite high for the VUV spectroscopy, however, the aim of these experiments was to synthesize residue from benzonitrile irradiation, so in order to synthesize sufficient residue and to obtain a good VUV spectrum of the residue the amount of sample used before irradiation resulted in a saturated VUV photoabsorption spectrum of benzonitrile at lower wavelengths (110 nm – 260 nm). However, in the higher wavelength region, 260 nm – 290 nm, due to the lower absorption cross section three bands at 264.5 nm, 271.6 nm and 279 nm were observed. After irradiation, these band positions and shape remain unchanged, however, the increase in baseline to higher absorption was an indication of absorption from newly synthesized molecule/residue synthesized by irradiation.

Using IR spectroscopy, in addition to the characteristic bands corresponding to the amorphous benzonitrile ice the appearance of a new band at 2133 cm$^{-1}$, that corresponds to the NC stretching [41], was clearly observed after irradiation. During warming of the ice IR spectra revealed phase change to create crystalline benzonitrile ice at 150 K. Further warming the ice to room temperature resulted in desorption of the benzonitrile but left a residue on the substrate. Such a residue was not observed in a benzonitrile ice that was warmed from 4 K to room temperature without irradiation.

The residue spectrum had absorption bands in the IR (Figure 1), quite similar to the benzonitrile characteristic absorption band suggesting the residue could be an aromatic derivative (containing nitrogen). In addition, the residue spectrum recorded in the VUV by warming the irradiated ice was also observed to contain the characteristic aromatic ring absorption at 190 nm and an extended absorption tail until 300 nm (Figure 1). Such a spectrum in the VUV is characteristic of aromatic derivatives [24]. Thus both spectroscopic techniques provide compelling evidence that the residue synthesized by VUV irradiation contains aromatic molecules. However, experimental evidence also suggests that the availability of carbon atoms in irradiated ices [42] may lead to the formation of carbon clusters or carbon only molecules [36].

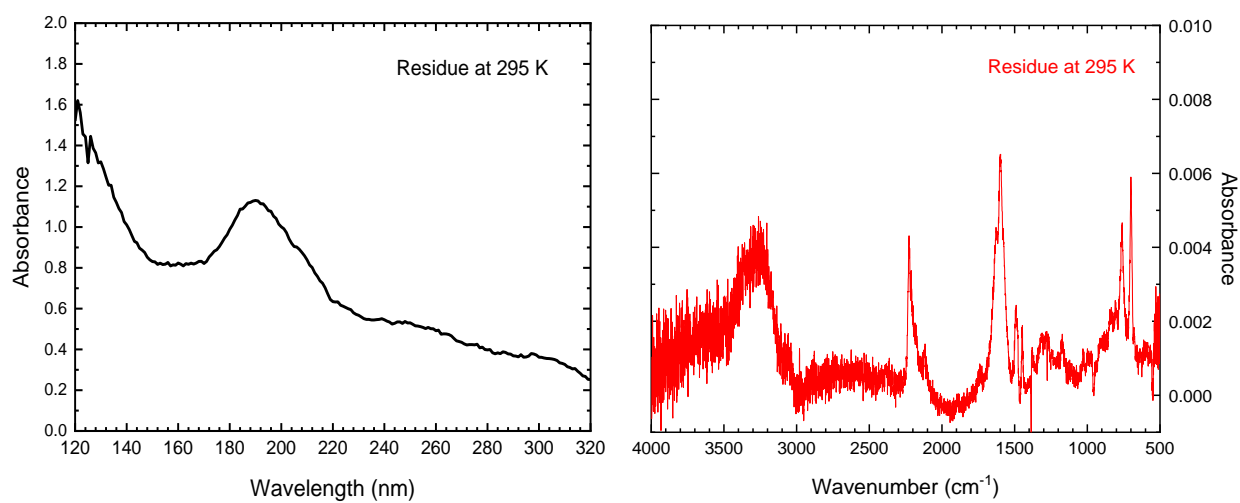

*Figure 1: Spectra of the residue, made from benzonitrile irradiation at 4 K and then warmed to room temperature, (left) in the VUV and (right) in IR wavelengths. The residue spectra were recorded at UHV conditions before decompressing the chamber to remove the windows.*

The appearance of characteristic features corresponding to non-volatile carbon / carbon-nitrogen clusters were not observed in VUV and IR spectra, but this could be due to lower concentrations in the residue. Therefore, in order to investigate the synthesis of non-volatile carbon only or carbon-nitrogen clusters from the ice irradiation, the sample was imaged using High Resolution - Transmission Electron Microscopy (HR-TEM), for which the residue obtained was scratched on to a Quantifoil TEM grid. Plasma cleaning, using a gas mixture of argon-oxygen, of the TEM grid was carried out before loading the grid for imaging.

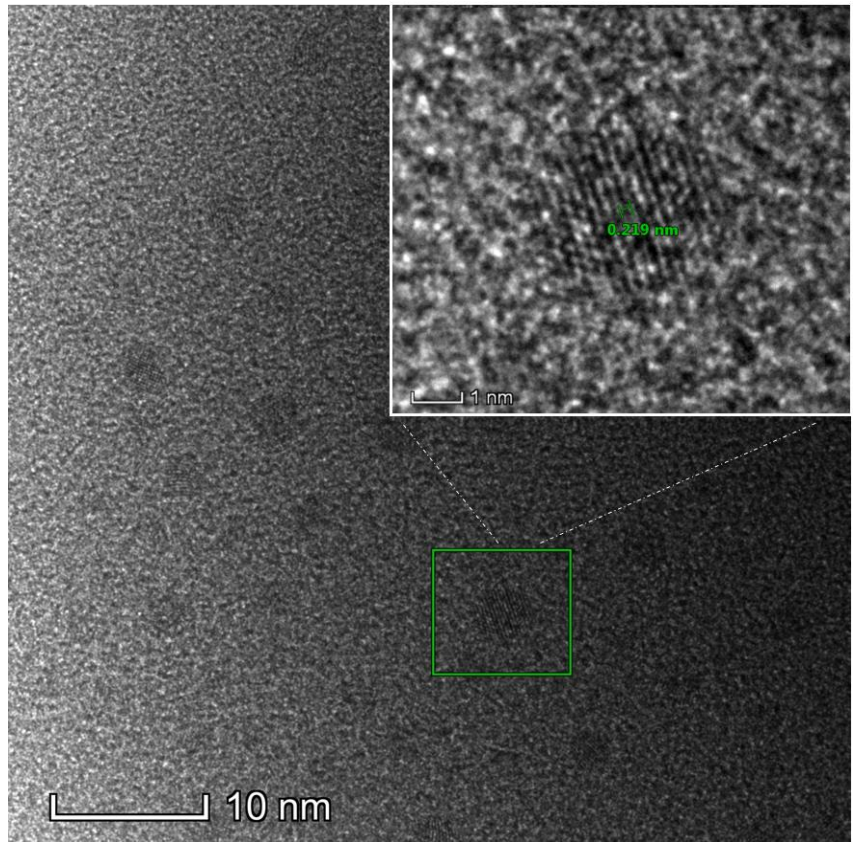

*Figure 2: HR-TEM image of the residue showing arrangements of atoms resembling a Quantum Dot.*

In the images obtained clusters of atoms arranged within the structure of a quantum dot, max length ~ 3 nm, from aligned atoms (from d spacing [43]) were observed. Such dots were seen in many places in the residue (Figure 2). Several nanosheets were also observed. Upon closer examination, on a scale of < 1 nm, this sheet was found to contain a hexagonal arrangement of atoms (Figure 3), a structure that which confirmed by the X-ray diffraction data (Figure 3). The energy dispersive spectrum showed carbon and nitrogen to be present in this sheet (Figure 3). From these observations we can conclude that within this residue there is evidence for the presence of nitrogen-doped graphene. In this case the quantum dot and N-doped graphene sheet were part of the other components present in a residue synthesized in laboratory analogue simulating ISM cold dust condition. The other components will be discussed in a separate publication.

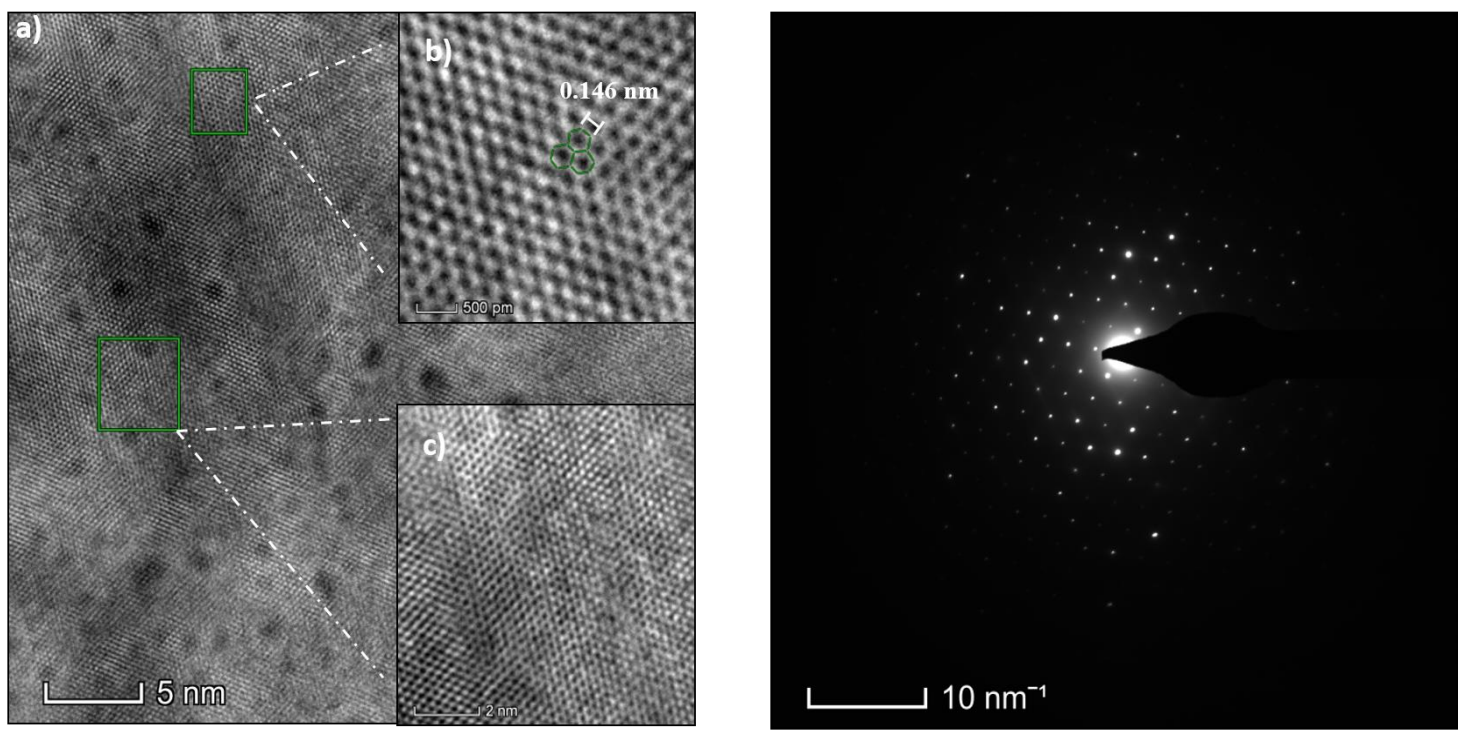
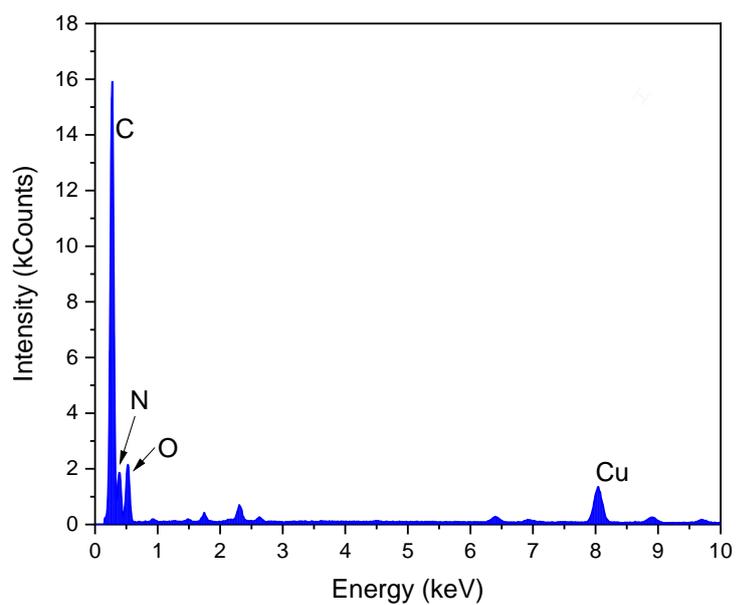

*Figure 3: HR-TEM image of the N-graphene sheet (left); (a) large sheet and the insets (b - c) shows the hexagonal pattern in two different parts of the large sheet. The corresponding diffraction pattern (right) and EDS (bottom).*

## 4. Discussion

A laboratory analogue experiment of the irradiation of an icy mantle made of the recently discovered benzonitrile molecule prepared under ISM conditions has revealed, for the first time, the formation of nanoscale structure analogous to a quantum dot and N-doped graphene sheet. This observation has important consequences for our understanding of the formation and abundance of PAH molecules in the ISM. A new and simple route to the formation of complex PAH structures is demonstrated by irradiation of a simple aromatic compound. The atoms of carbon and nitrogen released by photodissociation easily diffuses within the ice upon warming leading to the formation of nanostructures. The presence of graphene suggests that $C_{60}$ / $C_{70}$, via top-down model, and carbon nanotube, by rolling the sheet, may also form in irradiated aromatic ices. Graphene containing dust may be the source of $C_{24}$ reported by [28,29]. A variety of hydrocarbons and hydrocarbyls may result by etching graphene in hydrogen rich environments commensurate to conditions in the ISM [44]. While graphene in hydrogen rich environment can synthesize PAH molecules the presence of N-Graphene may synthesize Polycyclic Aromatic Nitrogen Heterocycles (PANHs), signatures of which may be present in spectral collection from the observations of different regions of the ISM clouds, using space and ground based observatories.

Titan's stratospheric polar clouds also contain benzene ices [45] and the chemical composition of surface ices on Titan are also known, from laboratory experiments [46], to harbour aromatic molecules such as benzene. Therefore, in a nitrogen rich environment where benzene ices are also present, benzonitrile may readily form [47]. Our experimental result showing N-graphene synthesis from benzonitrile ices therefore suggest that N-graphene may also be a component of Titan's icy clouds and surface ices. This prediction may be supported by CAPS ELS data collected during the T16 encounter of the Cassini spacecraft, at approximately 1000 km altitude [48], which with the observation of material with a few 1000's of Dalton. These high masses have been attributed to the presence of tholins [49] however N-graphene might co-exist with tholins in the chemically complex clouds of Titan. Therefore, adding the possible presence of N-graphene to atmospheric models of Titan may provide clues to those physico-chemical processes leading to formation of Titan's haze.

## 5. Conclusion

Irradiation of a 4 K benzonitrile ice film by 9 eV photons and subsequent warning to room temperature has demonstrated the formation of nanostructures including quantum dots and (Nitrogen-doped) graphene. This important observation shows that the bottom-up model to synthesize graphene is also feasible for a cold dust grain in the ISM, which can subsequently lead to the synthesis of larger PAH molecules in hydrogen rich environment. The search for graphene sheets in the ISM should therefore be made by looking at regions where benzonitrile has been identified. Astrochemical dust models should then consider the presence of graphene and how it many influences the productivity of PAH and other hydrocarbons.

In the Solar System, since benzonitrile is a likely component of the rich PAHs inventory present on Titan these results suggest that we can also expect the presence of N-graphene. Cassini flyby data should therefore be reviewed to look for any spectral signatures of N-graphene and Titan atmosphere models developed to explore synthetic routes of production in the Titan haze.

**Acknowledgements**

KKR, JKM, AB, PJ and BS thank the support from Physical Research Laboratory (Department of Space, Government of India). KKR, BNRS, AD and BS thank the support from Sir John Mason Academic Trust and Europlanet. Europlanet 2020 RI has received funding from the European Union's Horizon 2020 research and innovation programme under grant agreement No 654208. Europlanet 2024 RI has received funding from the European Union's Horizon 2020 research and innovation programme under grant agreement No 871149. KKR, BNRS, HH, DG, BMC, YJW, NJM and BS acknowledges the beamtime grant from NSRRC. BS acknowledges the INSPIRE Faculty Grant (IFA-11CH-11) under which the initial part of the work was carried out.

**References**

[1]     G. Strazzulla, G. A. Baratta, and M. E. Palumbo, Spectrochimica Acta Part A: Molecular and Biomolecular Spectroscopy **57**, 825 (2001).
[2]     M. H. Moore and B. Donn, The Astrophysical Journal **257**, L47 (1982).
[3]     L. J. Lanzerotti, W. L. Brown, and R. E. Johnson, in *Ices in the Solar System*, edited by J. Klinger *et al.* (Springer Netherlands, Dordrecht, 1985), pp. 317.
[4]     G. A. Baratta, V. Mennella, J. R. Brucato, L. Colangeli, G. Leto, M. E. Palumbo, and G. Strazzulla, Journal of Raman Spectroscopy **35**, 487 (2004).
[5]     R. Brunetto, M. A. Barucci, E. Dotto, and G. Strazzulla, The Astrophysical Journal **644**, 646 (2006).
[6]     P. Jenniskens, G. A. Baratta, A. Kouchi, M. S. de Groot, J. M. Greenberg, and G. Strazzulla, Astronomy and Astrophysics **273**, 583 (1993).
[7]     M. Nuevo, S. N. Milam, S. A. Sandford, B. T. De Gregorio, G. D. Cody, and A. L. D. Kilcoyne, Advances in Space Research **48**, 1126 (2011).
[8]     G. A. Baratta, M. M. Arena, G. Strazzulla, L. Colangeli, V. Mennella, and E. Bussoletti, Nuclear Instruments and Methods in Physics Research Section B: Beam Interactions with Materials and Atoms **116**, 195 (1996).
[9]     G. Ferini, G. A. Baratta, and M. E. Palumbo, Astronomy and Astrophysics **414**, 757 (2004).
[10]    Z. Kaňuchová, P. Boduch, A. Domaracka, M. E. Palumbo, H. Rothard, and G. Strazzulla, A&A **604** (2017).
[11]    M. H. Moore, R. L. Hudson, and R. W. Carlson, Icarus **189**, 409 (2007).
[12]    D. Sicilia, S. Ioppolo, T. Vindigni, G. A. Baratta, and M. E. Palumbo, A&A **543** (2012).
[13]    O. Gomis and G. Strazzulla, Icarus **194**, 146 (2008).

[14] G. Strazzulla, M. Garozzo, and O. Gomis, Advances in Space Research **43**, 1442 (2009).
[15] R. E. Johnson, J. F. Cooper, L. J. Lanzerotti, and G. Strazzulla, Astronomy and Astrophysics **187**, 889 (1987).
[16] G. Strazzulla, G. A. Baratta, R. E. Johnson, and B. Donn, Icarus **91**, 101 (1991).
[17] O. Poch *et al.*, Science **367**, eaaw7462 (2020).
[18] K. Altwegg *et al.*, Nature Astronomy **4**, 533 (2020).
[19] J. P. Dworkin, D. W. Deamer, S. A. Sandford, and L. J. Allamandola, Proceedings of the National Academy of Sciences **98**, 815 (2001).
[20] A. Potapov, C. Jäger, and T. Henning, Physical Review Letters **124**, 221103 (2020).
[21] L. Calcagno, G. Foti, L. Torrisi, and G. Strazzulla, Icarus **63**, 31 (1985).
[22] G. Strazzulla, L. Calcagno, and G. Foti, Monthly Notices of the Royal Astronomical Society **204**, 59P (1983).
[23] G. Foti, L. Calcagno, K. L. Sheng, and G. Strazzulla, Nature **310**, 126 (1984).
[24] K. K. Rahul *et al.*, Spectrochimica Acta Part A: Molecular and Biomolecular Spectroscopy **231**, 117797 (2020).
[25] M. P. Callahan, P. A. Gerakines, M. G. Martin, Z. Peeters, and R. L. Hudson, Icarus **226**, 1201 (2013).
[26] S. Kwok and Y. Zhang, Nature **479**, 80 (2011).
[27] S. Kwok and Y. Zhang, The Astrophysical Journal **771**, 5 (2013).
[28] D. A. García-Hernández, N. K. Rao, and D. L. Lambert, The Astrophysical Journal **729**, 126 (2011).
[29] D. A. García-Hernández, E. Villaver, P. García-Lario, J. A. Acosta-Pulido, A. Manchado, L. Stanghellini, R. A. Shaw, and F. Cataldo, The Astrophysical Journal **760**, 107 (2012).
[30] Q. Li, A. Li, and B. W. Jiang, Monthly Notices of the Royal Astronomical Society **490**, 3875 (2019).
[31] J. Cami, J. Bernard-Salas, E. Peeters, and S. E. Malek, Science **329**, 1180 (2010).
[32] H. W. Kroto, J. R. Heath, S. C. O'Brien, R. F. Curl, and R. E. Smalley, Nature **318**, 162 (1985).
[33] P. Merino *et al.*, Nature Communications **5**, 3054 (2014).
[34] A. Chuvilin, U. Kaiser, E. Bichoutskaia, N. A. Besley, and A. N. Khlobystov, Nature Chemistry **2**, 450 (2010).
[35] O. Berné and A. G. G. M. Tielens, Proceedings of the National Academy of Sciences **109**, 401 (2012).
[36] M.-Y. Lin, J.-I. Lo, H.-C. Lu, S.-L. Chou, Y.-C. Peng, B.-M. Cheng, and J. F. Ogilvie, The Journal of Physical Chemistry A **118**, 3438 (2014).
[37] B. A. McGuire, A. M. Burkhardt, S. Kalenskii, C. N. Shingledecker, A. J. Remijan, E. Herbst, and M. C. McCarthy, Science **359**, 202 (2018).
[38] I. R. Cooke, D. Gupta, J. P. Messinger, and I. R. Sims, The Astrophysical Journal **891**, L41 (2020).
[39] H.-C. Lu, H.-K. Chen, B.-M. Cheng, Y.-P. Kuo, and J. F. Ogilvie, Journal of Physics B: Atomic, Molecular and Optical Physics **38**, 3693 (2005).
[40] H.-C. Lu, H.-K. Chen, B.-M. Cheng, and J. F. Ogilvie, Spectrochimica Acta Part A: Molecular and Biomolecular Spectroscopy **71**, 1485 (2008).
[41] M. Accolla, G. Pellegrino, G. A. Baratta, G. G. Condorelli, G. Fedoseev, C. Scirè, M. E. Palumbo, and G. Strazzulla, A&A **620**, A123 (2018).
[42] N. E. Sie, G. M. M. Caro, Z. H. Huang, R. Martín-Doménech, A. Fuente, and Y. J. Chen, The Astrophysical Journal **874**, 35 (2019).
[43] P. Tian, L. Tang, K. S. Teng, and S. P. Lau, Materials Today Chemistry **10**, 221 (2018).
[44] J. I. Martínez, J. A. Martín-Gago, J. Cernicharo, and P. L. de Andres, The Journal of Physical Chemistry C **118**, 26882 (2014).
[45] S. Vinatier, B. Schmitt, B. Bézard, P. Rannou, C. Dauphin, R. de Kok, D. E. Jennings, and F. M. Flasar, Icarus **310**, 89 (2018).
[46] M. J. Abplanalp, R. Frigge, and R. I. Kaiser, Science Advances **5**, eaaw5841 (2019).

[47]     J. C. Loison, M. Dobrijevic, and K. M. Hickson, Icarus **329**, 55 (2019).
[48]     J. H. Waite, D. T. Young, T. E. Cravens, A. J. Coates, F. J. Crary, B. Magee, and J. Westlake, Science **316**, 870 (2007).
[49]     C. Sagan and B. N. Khare, Nature **277**, 102 (1979).